# The cosmic ray test of MRPCs for the BESIII ETOF upgrade[*]

Wang Xiao-Zhuang(王小状)[1,2,3;1)] Heng Yue-Kun(衡月昆)[2,3;2)] Wu Zhi(吴智)[2,3] Li Cheng(李澄)[1,2]
Sun Yong-Jie(孙勇杰)[1,2] Dai Hong-Liang(代洪亮)[2,3] Sun Sheng-Sen(孙胜森)[2,3] Yang Rong-Xing(杨荣星)[1,2]
Cao Ping(曹平)[1,2] Zhang Jie(张杰)[2,3] Wang Yun(汪昀)[1,2] Sun Wei-Jia(孙维佳)[1,2] Wang Si-Yu(王思宇)[1,2]
Ji Xiao-Lu(季筱璐)[2,3] Zhao Jin-Zhou(赵金周)[2,3] Gong Wen-Xuan(龚文煊)[2,3] Ye Mei(叶梅)[2,3]
Ma Xiao-Yan(马骁妍)[2,3] Chen Ming-Ming(陈明明)[2,3] Xu Mei-hang(徐美杭)[2,3] LuoXiao-Lan(罗小兰)[2,3]
Zhu Ke-Jun(朱科军)[2,3] Liu Zhen-An (刘振安)[2,3] Jiang Xiao-Shan(江晓山)[2,3]

1 Department of Modern Physics, University of Science and Technology of China, Hefei 230026, China

2 State Key Laboratory of Particle Detection and Electronics, Beijing 100049, China

3 Institute of High Energy Physics, Chinese Academy of Sciences, Beijing 100049, China

**Abstract:** In order to improve the particle identification capability of the Beijing Spectrometer III (BESIII), it is proposed to upgrade the current endcap time-of-flight (ETOF) detector with multi-gap resistive plate chamber (MRPC) technology. Aiming at extending ETOF overall time resolution better than 100ps, the whole system including MRPC detectors, new-designed Front End Electronics (FEE), CLOCK module, fast control boards and time to digital modules (TDIG), was built up and operated online 3 months under the cosmic ray. The main purposes of cosmic ray test are checking the detectors' construction quality, testing the joint operation of all instruments and guaranteeing the performance of the system. The results imply MRPC time resolution better than 100ps, efficiency is about 98% and the noise rate of strip is lower than 1Hz/(s cm$^2$) at normal threshold range, the details are discussed and analyzed specifically in this paper. The test indicates that the whole ETOF system would work well and satisfy the requirements of upgrade.

## 1 Introduction

The Beijing Spectrometer III(BESIII )[1,2] is a high precision general-purpose detector designed for high luminosity e$^+$e$^-$ collisions in the τ-charm energy region at the Beijing Electron and Positron Collider II(BEPCII) [3-5]. In high energy collision experiments particle identification (PID) play a crucial role, especially for hadrons analysis in τ-charm physics. Currently at BESIII the PID detector are dominated by time-of-fight(TOF) detector which consists of barrel TOF(BTOF)[6] and Endcap TOF(ETOF)[7] that is at the region 0.83<cosθ <0.90 ( θ is the polar angle as shown in Fig. 1).The current ETOF consists of fast scintillators (BC204) and fine-mesh photomultiplier tubes (Hamamatsu R5924) [8]. The time resolution measured by ETOF detector is 138ps for π(1GeV/c) and the momentum range of K/π separation (2σ) is limited to 1.1GeV/c [6,7], which cannot satisfy the higher precision requirement of physics analysis. One of the reasons worsening ETOF resolution is scattering particle. The secondary particles mainly photons created from the multiple scatterings upon the materials between MDC end-cap and E-TOF leads to a high multi-hits rate (per channel), especially for electron events (~71.5%)[9]. These multi-hits events distort the shape and amplitude of the output signals, make the offline calibration difficult and degrade the time resolution.

---

[*] This work is supported by the National Natural Science Foundation of China (Grant No. *10979003*) .

1) E-mail: xzwang@ihep.ac.cn

2) E-mail: hengyk@ihep.ac.cn

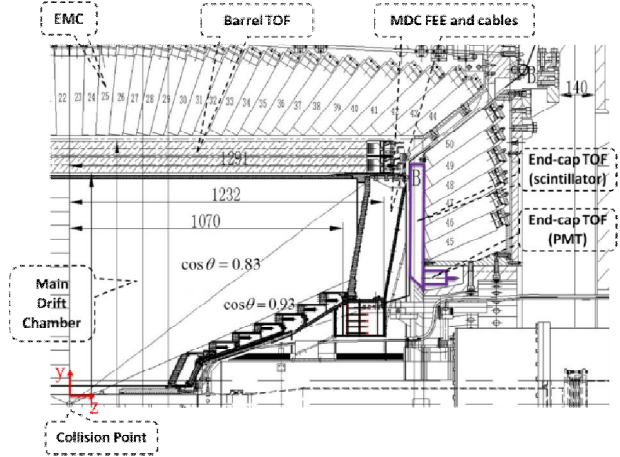
Fig.1.Schematicsof E-TOF location of BESIII

The multi-gap resistive plate chamber (MRPC)[10] as a new type of gaseous detector, with good time resolution, high detection efficiency, relatively low cost and insensitive to neutral particles, has been widely used as TOF detector in many experiments currently, such as ALICE at LHC[11,12], STAR at RHIC[13,14]. In BESIII a proposal[15] was approved in 2012 to upgrade the current ETOF with the MRPC technology, aiming at improving the overall time resolution better than 100ps for π, so that the K/π separation (2σ) momentum range can be extended to 1.4GeV/c at BESIII. Considering the non-intrinsic contributions, such as the beam size, Z uncertainty etc., the resolution of ETOF system is required at 80~120ps for 0.8GeV/c momentum π while intrinsic resolution of MRPC is required to be better than 80ps.

In this paper, the cosmic ray test of the whole ETOF upgrading system is presented. In the following sections, the MRPC modules, the readout electronics, the high voltage system, the gas system, and data acquisition system are described. The results of system test and analysis method will be discussed in detail. At the last section a conclusion will be given.

## 2 ETOF upgrade Project

### 2.1 Design of BESIII ETOF

In project design [15], there are total 72 MRPCs in ETOF at BESIII, and each endcap has 36 overlapping MRPCs, as shown in Fig.2. They are separated into 2 tiers with 18 MRPC modules. The effective inner radius of the ring is 478 mm while the outer radius is 822 mm.

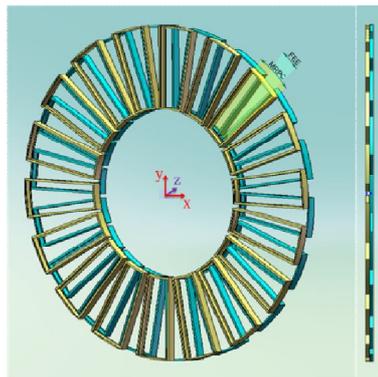
Fig.2.Structure each endcap of BESIII E-TOF MRPC ring

### 2.2 Structure of ETOF MRPC

The double-end readout MRPC module has 12 readout cells with the lengths ranging from 9.1 cm to 14.1 cm. The width of each strip is 2.4cm and interval between strips is 3 mm, as shown in Fig.3. Comparing with single-end readout strip MRPC [16,17], the time resolution of double-end readout is less affected by the hit position with the mean time from the two strips. And the time difference from both ends can also provides some position information along the strips.

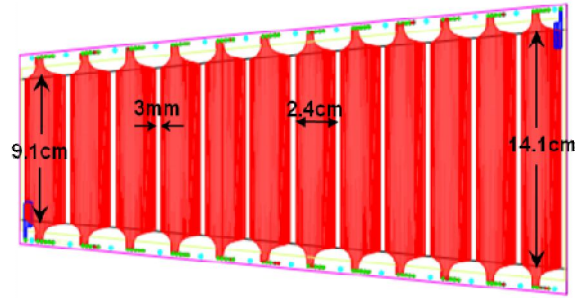

Fig.3. The layouts of Printed Circuit Board (PCB) of MRPC from top view

The structure of MRPC is shown in Fig.4. There are 12 gas gaps arranged in two MRPC stacks. Each gap thickness is 0.22 mm which is defined by the nylon fishing line between the glass plates whose volume resistance keep at $10^{13}\Omega$ cm. The thicknesses of glasses are 0.4 mm and 0.55 mm for the inner and outer glass, respectively. The outer surfaces of the outermost glass in each stack are coated with graphite tapes, which serve as high voltage electrodes. The surface resistivity of the graphite tape is about 200 k$\Omega$/□ .Two pieces of 3 mm thick honeycomb-board are attached to the outer surfaces of the detector to reduce structural deformations. The MRPC prototype module is placed in a gas-tight aluminium box whose total thickness is 2.5 mm (~$0.028X_0$), flushed with a standard gas mixture (90% Freon+5% $SF_6$+5% iso-$C_4H_{10}$).

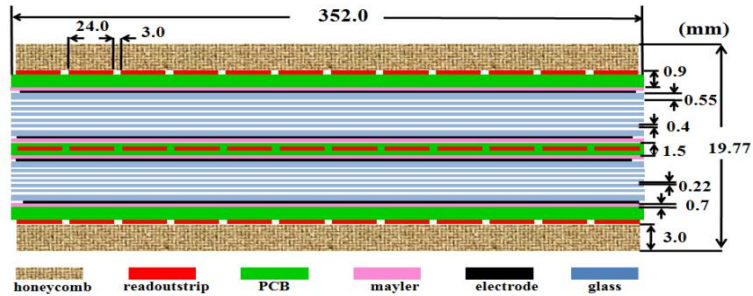

Fig.4. The schematic of MRPC from cross-section view

## 2.3 Readout electronics and data acquisition system

The electronics system equipped for MRPCs, is mainly composed of FEEs, TDIGs, fast control module and CLOCK module in NIM crates which are together with the data acquisition (DAQ) system. The FEE is used the NINO chip developed by the ALICE-TOF group [18]. Each FEE module features 24 differential input channels and outputs correspondent LVDS signal with the signal charge encoded in its width. The timing accuracy (RMS) can be better than 15 ps for each channel when the input charge is larger than 100 fC [19]. The FEE board is fixed on the surface of the aluminium gas box which contains the MRPC module in order to save space of connectors and cables. A flexible printed circuit is designed to connect the MRPC module output. The connector (VRDPC-68-01-M-RA) with 50 pins and the shielded differential cable (VPSTP-24-5000-01 from SAMTEC) are used to connect the FEE and the TDIG [20], which are aiming at reducing the time jitter from signal transmission and ensuring the signal quality.

The Calibration Threshold Test Power (CTTP) module play a important role in electronics system, which is housed in NIM crate, provides power, threshold and test signals to the FEE. It also receives the OR differential signals from the FEE and produces fast logic signals. The hit signals are used by the TOF trigger subsystem.

The 9U VME crate consists of the VME controller of PowerPC, the readout control module of ROC, the clock production module, and the TDIG module. The TDIG modules, relying on the ASIC HPTDC chip developed by the microelectronics group at CERN[21]. HPTDC measures the absolute timing on both leading edge and trailing edge. All the signals are recorded in each channel continuously in its typical time-stamp style measurement shown as in Figure.5, the edges of any signals are digitalized one after another with respect to the clock edge. The time stamps are then saved into the buffer. After receiving and digitizing the signals, data packing and uploading in TDIG modules with predefined format are operated by the data acquisition (DAQ) system via the VME bus.

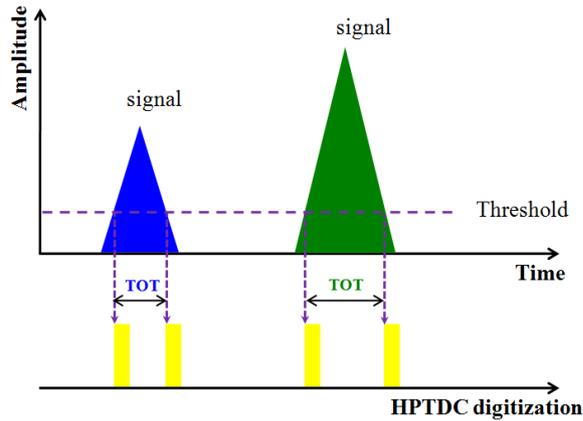

Fig.5. The diagram of signals consequence in HPTDC digitization

There are two working modes for the electronics system, namely data taking and calibration, which are determined by the ROC module and the latter mode is mainly used to test if the system works properly. At the mode of data taking, the ROC receives a series of control signals, such as "clock", "trigger" and sends them to the module of TDIG to start the system, while at the calibration mode, the ROC generates these signals by itself. The ROC module sends signals to CTTP to generate the test signals and to TDIG to control the time measurements. Upon the completion for measurements, the ROC module will produce an interrupt signal on VME bus, and then data acquisition system will read out the data from the TDIG module and deliver them to the PC.

The data acquisition system is used the same as that of the BESIII experiment, adopting the techniques of multi-level buffering, parallel processing, high-speed VME readout and network transmission. Also, the running status such as the event rate, noise level, the time spectrum can be monitored by a program at the PC to give a real-time information for the whole system. The resolution of whole electronics system is about ~25ps after an integral non-linearity (INL) compensation [10]. The schematic of the readout electronics system for the ETOF upgrade is shown in Fig. 5.and Fig.6.

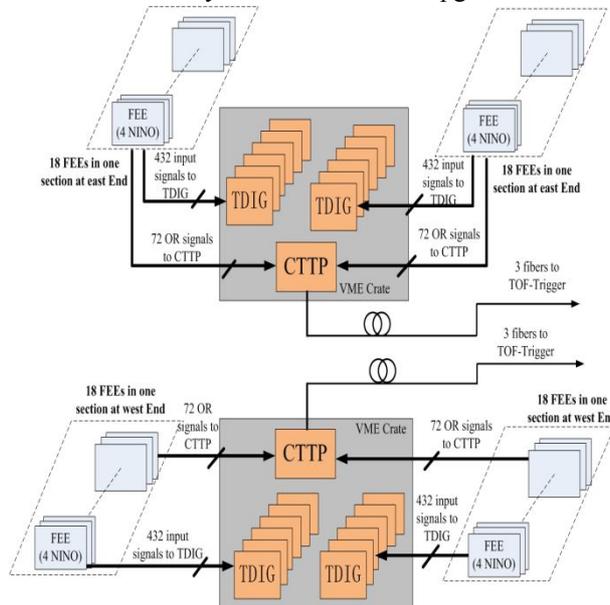

Fig.5.The schematic of readout electronics system

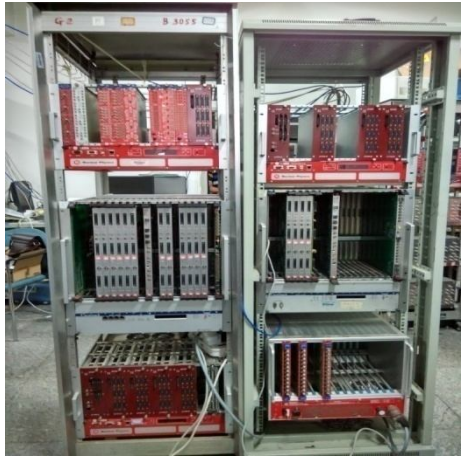

Fig.6.The readout electronics system

## 2.3 Gas system and high voltage system

The gas system consists of the gas supply bottle, mass and flow monitors and gas tanks. The component of working gas is the standard 90%Freon + 5%$SF_6$ + 5%iso-$C_4H_{10}$ for the test. The gas ratios are controlled by mass flow meters, which is flew into a buffer tank. The flow flux supplied to the MRPC modules is fixed at a rate of 400ml/min. In order to avoid the effect of moisture, copper pipes are chosen to connect between gas tanks and MRPC modules. Plastic pipes are chosen for the output of the gas. Each three MRPC modules are connected serially in the gas circuit.

The high voltage system contains nine pairs of positive and negative channels, consists of the high voltage (HV) power supply, the HV distribution crate and flexible connection cables. Each pair of channels are split by the HV distribution crate to supply four modules and controlled by the BESIII slow control system.

# 3 The Cosmic ray test
## 3.1 The system layout

The overlapped MRPC modules are stacked on 4 semicircle platforms as shown in Fig.7. To trigger the passing particles, 18pairs of trapezium scintillator counters, the active areas of which are slightly same as the MRPC active area, was placed above and under the MRPC module separately. The logic diagram of the system setup is shown in Fig.8.

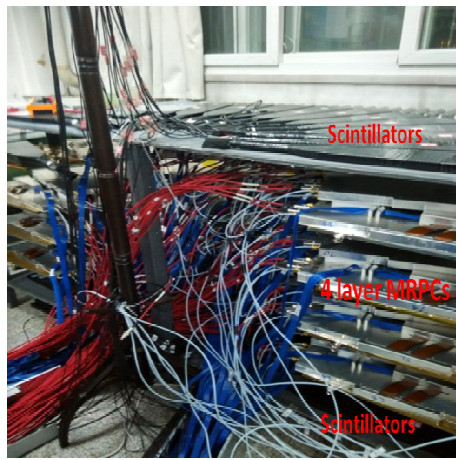

Fig.7. The planting setup of MRPCs and scintillators

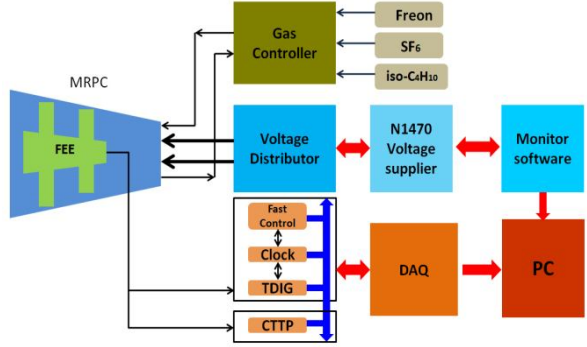

Fig.8. The logic diagram of system setup

### 3.2 Data analysis

In this section, we discuss the data analysis. For the convenience of data analysis, we select same strip pad of each vertical four MRPC modules in each cosmic ray event. In Fig.9, it shows one cosmic ray event passing through the same number strip of four modules. We define $t_{ia}$, $T_i$, and $Tr_i$ to easily and effectively emphasized our analysis process bellow. $t_{ia}$ is used to describe the arrival time which $i(i=1,2,3,4)$ represents the different module and $a(a=1,2)$ represents the two ends of strip pad. $T_i = (t_{i1}+t_{i2})/2$ can be regarded as average hit-time of the strip. While reference time $Tr_i = (T_j+T_k+T_l)/3$, which represents the other 3 strips' hit-time of $T_i$ (note that $Tr_i$ is irrelevant to $TOT_i$ since they are not from same MRPC). Resultantly 4 $T_i$ - $Tr_i$ distributions can be obtained and preliminary resolutions of them could be achieved actually. Considering Time-Amplitude effect, each time $T_i$ is dependant with the amplitude of the signal, which is substituted by $TOT_i$ in our electronics readout system, inevitably $T_i$-$Tr_i$ must get rid of the jitter from the every $TOT_i$. Generally this effect is corrected with so-called T-A slewing method. It can be explained briefly bellow in our analysis example: First we fit $T_i$-$Tr_i$ versus $TOT_i$ relationship with empirical function, Here is MRPC's in Equation(1), also indicated from Fig.11:

$$f(x)=p_0+\frac{p_1}{\sqrt{x}}+\frac{p_2}{x}+\frac{p_3}{x\sqrt{x}}+\frac{p_4}{x^2} \tag{1}$$

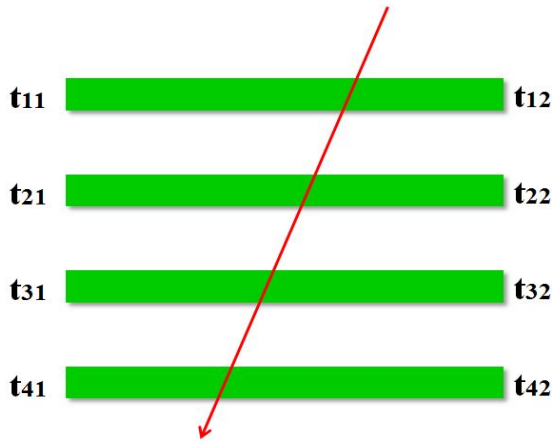

Fig.9. The model of strips group hit by a cosmic ray event

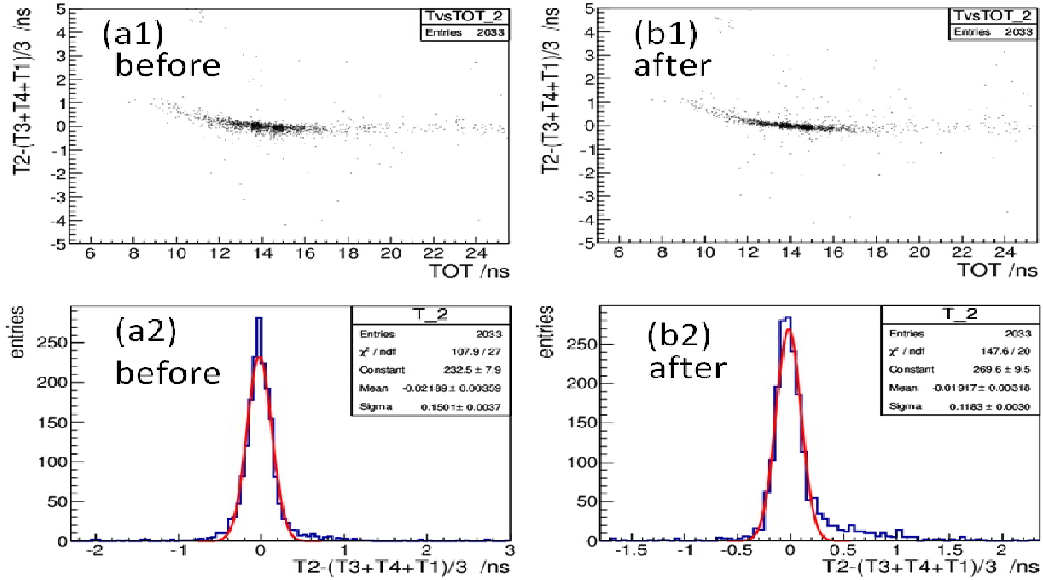

Fig.10. The compare example of reference time $Tr_i$ befeore and after 3 times iterations and slewings

Secondly, using the fit function corrected time component $T_i$ in $Tr_i$ by $T_i - f(TOT_i)$ of each strip instead of preliminary time in $Tr_i$, which means $Tr_i$ becomes less dependent with the $TOT_i$ effect. A example is emphasized in Fig.10. We can repeat the slewing process n(n=1,2,…) iterations until the $Tr^{(n)}_i$ can be treated irrelevant to $TOT_i$. Then we do the last slewing in $T_i - Tr^{(n)}_i$ versus $TOT_i$, ultimately 4 resolutions distributions were all created independent with TOT which will be most exactest. According to Fig.11 and Fig.12, it can be illustrated by the example of final resolution distribution. Generally we find that almost after 3 times iterations the results become stable and it is similar to those after more iterations. This method will be very helpful for the performance test in the mass production since it can get the test results of several MRPC modules simultaneously.

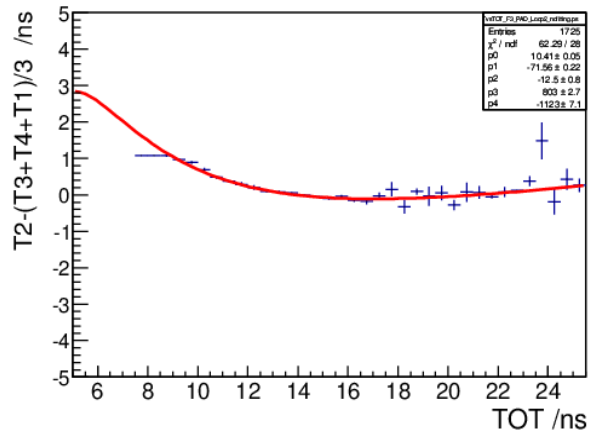

Fig.11. A T-TOT fitted fine with empirical function

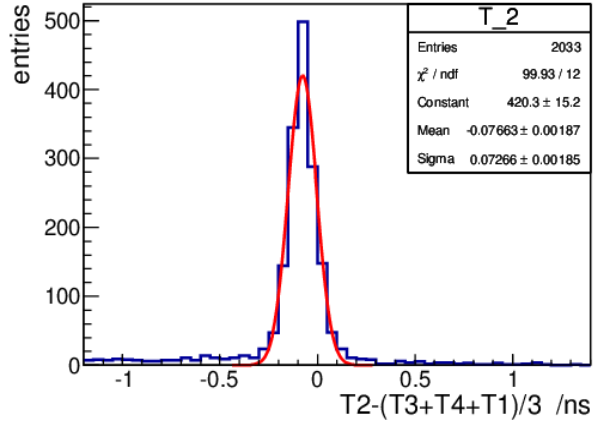

Fig.12. A example of T$_i$-Tr$_i$ distribution after 3 times iterations and slewing correction

At last after several iterations and slewing corrections, four sigmas of distributions T$_i$-Tr$_i$ could be measured. And here define $\sigma_{mi}$ ($i$= 1, 2, 3, 4) as measured resolution which eventually is most accurate from each last fitting function, while $\sigma_i$ represents the true intrinsic resolution of each MRPC. Finally resolutions of four strips group can be described in Equation (2).

$$\begin{cases} \sigma_{m1}^2 = \sigma_1^2 + \left(\sigma_2^2 + \sigma_3^2 + \sigma_4^2\right)/9 \\ \sigma_{m2}^2 = \sigma_2^2 + \left(\sigma_3^2 + \sigma_4^2 + \sigma_1^2\right)/9 \\ \sigma_{m3}^2 = \sigma_3^2 + \left(\sigma_4^2 + \sigma_1^2 + \sigma_2^2\right)/9 \\ \sigma_{m4}^2 = \sigma_4^2 + \left(\sigma_1^2 + \sigma_2^2 + \sigma_3^2\right)/9 \end{cases} \quad (2)$$

After solving this matrix equation, the intrinsic time resolutions matrix of each can be obtained by the formula (3):

$$\begin{pmatrix} \sigma_1^2 \\ \sigma_2^2 \\ \sigma_3^2 \\ \sigma_4^2 \end{pmatrix} = \begin{pmatrix} 33/32 & -3/32 & -3/32 & -3/32 \\ -3/32 & 33/32 & -3/32 & -3/32 \\ -3/32 & -3/32 & 33/32 & -3/32 \\ -3/32 & -3/32 & -3/32 & 33/32 \end{pmatrix} \begin{pmatrix} \sigma_{m1}^2 \\ \sigma_{m2}^2 \\ \sigma_{m3}^2 \\ \sigma_{m4}^2 \end{pmatrix} \quad (3)$$

### 3.3 Test results

Considering the jitter of electronics[10] (~25ps) the intrinsic resolution of 12 strips in MRPCs results are shown in Fig.14. The average intrinsic resolution of MRPC strip is about 60ps. The resolutions of shorter strips and longest strip are a bit worse than these middle strips probably due to detection edge effect and low statistics, and also the uncertainty from momentum and flight between strips of cosmic ray are both possible factors which deteriorate resolution effectively.

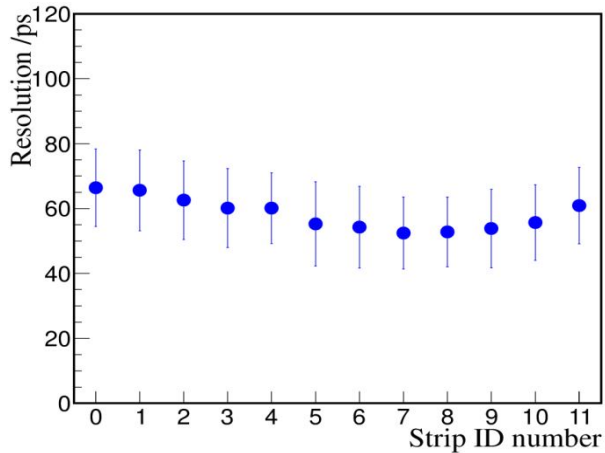

Fig.14. The resolutions versus strip ID number

The detecting efficiency is determined by the ratio of the number of hits of two-end in one pad to the number of tracks passing through the pad. Because there is no tracker in the system, we use MRPC1 and MRPC4 as the reference one. When studying the efficiency of strip pad middle MRPCs, the same pads of MRPC1 and MRPC4 are required to have hits in both ends, and the TOT of this pad should be larger than that of the adjacent ones. According to Fig.15, the strip efficiency is about ~98%, which was more than the requirement of ETOF upgrade.

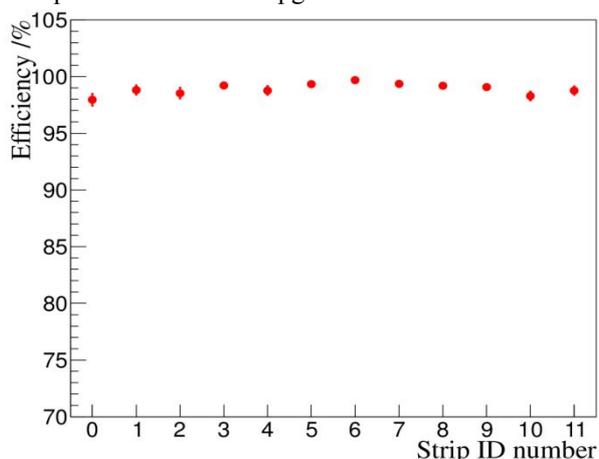

Fig.15.The detecting efficiency versus strip ID number

The noise rate of strips with every electronic channel was also studied, a signal which was independent of the cosmic ray was sent to the electronics as random "trigger". With the different threshold values setting into the NINO chip, the average noise rate of MRPC strips reduces and keeps steady below 1Hz/scm$^2$ as shown in Fig.16. The figure indicates that in normal detector working threshold range the average noise rate keeps at low level and our system is absolutely confident with data reception and storage.

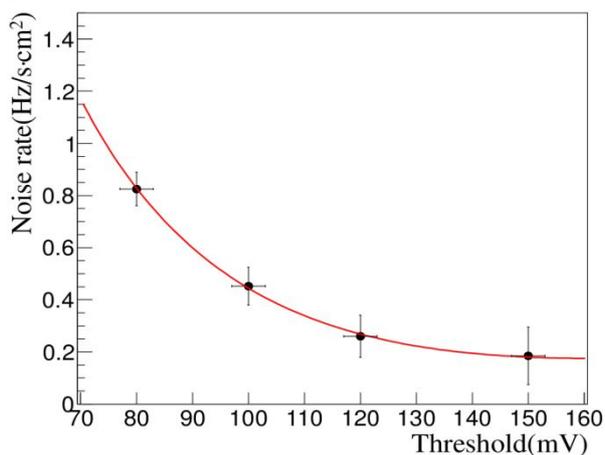

Fig.16. The average noise ratio of channels in different threshold

## 4 Conclusion

The upgrade of the BESIII E-TOF with MRPC technology has been approved and is going on smoothly. The cosmic ray test system consisted of MRPCs and electronics system designed for the upgrade project was built and tested successfully. The intrinsic time resolution of MRPC module is about 60ps under ±7000 HV, the detecting efficiency is about 98% which is above requirement value 96% and the average noise rate of strip is lower than 1Hz/s cm$^2$ at working threshold range. The results indicate the whole system of ETOF could work steady under cosmic ray test.

### Acknowledgments

*This work is supported by the National Natural Science Foundation of China under Grant No. 10979003,11275196.*